\documentclass[a4paper,12pt]{cernart} 
\usepackage[dvips]{graphicx}
\begin{document}
\newcommand{\epr}{\varepsilon^\prime\!/\varepsilon}
\newcommand{\repr}{\mathrm{Re}\,(\epr)}
\newcommand{\ko}{K^0}
\newcommand{\kobar}{\bar{K^0}}
\newcommand{\kl}{K_L}
\newcommand{\ks}{K_S}
\newcommand{\pipin}{\pi^0\,\pi^0\,}
\newcommand{\pipic}{\pi^+\pi^-}
\newcommand{\kethree}{K_{e3}}
\newcommand{\kmuthree}{K_{\mu3}}
\newcommand{\ptprime}{p_t{}^\prime}
\newcommand{\asp}{\mathcal{A}}
\begin{titlepage}
\vspace{1.2cm}
\title{MEASUREMENT OF DIRECT CP VIOLATION\\
WITH THE EXPERIMENT NA48 AT CERN}
\begin{Authlist}
Sandro Palestini \Aref{a}
\Instfoot{a1}{CERN, 1211 Geneva 23, Switzerland \Aref{b}}
\end{Authlist}
\vspace{2.5cm}
\begin{abstract}\noindent
Direct CP violation is studied in two pion decays 
of the neutral kaon.
Using data collected in the first beam period in 1997, 
the result for the parameter $\repr$ is
$(18.5 \pm 4.5 \mathrm{(stat)} \pm 5.8 \mathrm{(syst)})\times10^{-4}$.
\end{abstract}
\vspace{3cm}
\conference{Talk presented at the EPS--HEP 99 Conference\\
Tampere, Finland, July 15--21, 1999}
\Anotfoot{a}{On behalf of the NA48 Collaboration:
Cagliari, Cam\-bridge, CERN, Dubna, Edinburgh, Ferrara, Firenze, Mainz, Orsay,
Perugia, Pisa, Saclay, Siegen, Torino, Vienna, Warsaw.}
\Anotfoot{b}{Permanent address: INFN, sez.\ Torino, 10125 Torino, Italy.}
\end{titlepage}
\section{Introduction}
Two pion decays of $\kl$ \cite{ref:cpv} established the 
violation of the symmetry CP.
The main effect is due to a small component 
of CP = $+\,1$ eigenstate in $\kl$,  
which decays into two pions in a way similar to 
$\ks$. Direct CP violation is found in the deviation from unity 
of the double ratio of decay rates:
\begin{eqnarray}
 R & = &  
   \frac{\Gamma(\kl \rightarrow \pipin\,) /\, \Gamma(\ks \rightarrow \pipin\,)}
        {\Gamma(\kl \rightarrow \pipic) /\, \Gamma(\ks \rightarrow \pipic)}
\nonumber \\
   & \simeq & 1-6\times \repr \:.\nonumber
\end{eqnarray}
First evidence for non--vanishing $\repr$ \cite{ref:na31} was not supported 
by a different experiment \cite{ref:e731}, while a recent result 
\cite{ref:ktev} confirms the effect. 
 
The Standard Model provides a natural mech\-a\-nism for direct CP violation, 
but current pre\-dic\-tions \cite{ref:buchalla,ref:computations} 
are affected by significant computational uncertainties.

The experiment NA48 has been designed for an accurate measurement of $\repr $, 
using an approach with significant improvements 
over the techniques used previously \cite{ref:na31,ref:e731}. 
The first results, from data collected in 1997, 
are presented here.

\section{Experimental technique}
In order to minimize the sensitivity to detector efficiency,
to variations in beam intensity, and to accidental activity,
the experiment \cite{ref:na48}  
is designed to collect data si\-mul\-ta\-ne\-ous\-ly in the 
four channels $\kl$, $\ks\rightarrow \pipin$, $\pipic$, with the kaon energy 
in the interval 70--170 GeV. Two neutral beams are used: 
the $\kl$ beam is produced 126~m 
\begin{figure}[b]
\begin{center}
\includegraphics*[scale=0.325]{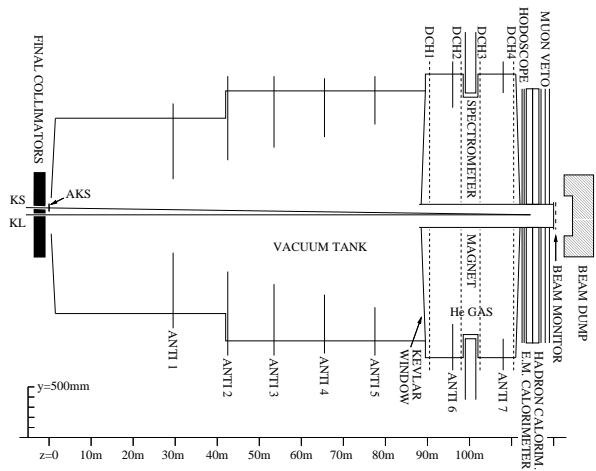}  
\caption {Layout of the main detector components.}
\label {fig:detector}
\end{center}
\end {figure} 
\begin{figure}[ht]
\begin{center}
\includegraphics*[scale=0.425]{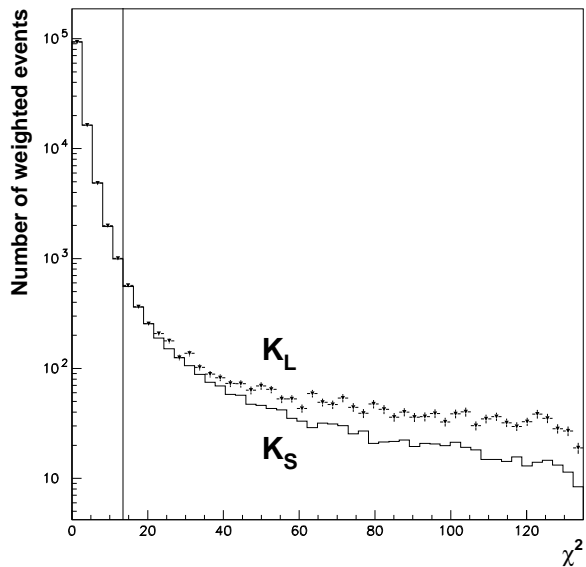} 
\caption
{Distribution of $\chi^2$ for $\kl$ and 
$\ks \rightarrow \pipin$, normalized in the first bin.
The excess of $K_L$ candidates in the region
$\chi^2$ $>$ $36$ is used to compute the background due to
3$\pi^0$ in the signal region ($\chi^2$ $<$ 13.5).}
\label {fig:bkgn}
\end{center}
\end {figure} 
(cor\-re\-spond\-ing on average to 21 $\ks$ lifetimes $\tau_S$) 
upstream of the nominal decay region. The $\ks$ beam is produced 6~m 
(one $\tau_S$) upstream of the decays region. The two beams, which are 68~mm 
apart as they pass the final collimator, converge and cross at the 
position of the electromagnetic (e.m.) calorimeter, 115~m downstream.
Figure~\ref{fig:detector} shows the layout of the decay region and the main
detector components. In order to minimize the difference in acceptance 
between $\kl$ and $\ks$, only decays occurring in the upstream part are 
used, requiring $0\!<\!\tau\!<\!3.5 \:\tau_S$, which corresponds to 
$0\!<\!z\!<\!21$~m on average.  

The identification $\kl$ vs.\  $\ks$ is done looking for a time coincidence 
between a kaon decay measured in the main detector, and the detection of 
a proton in the beam directed to the 
$\ks$ production target. This is done by means of
a finely segmented scin\-til\-lation counter (tagger).
\begin {figure}[p]
\begin{center}
\includegraphics*[scale=0.425]{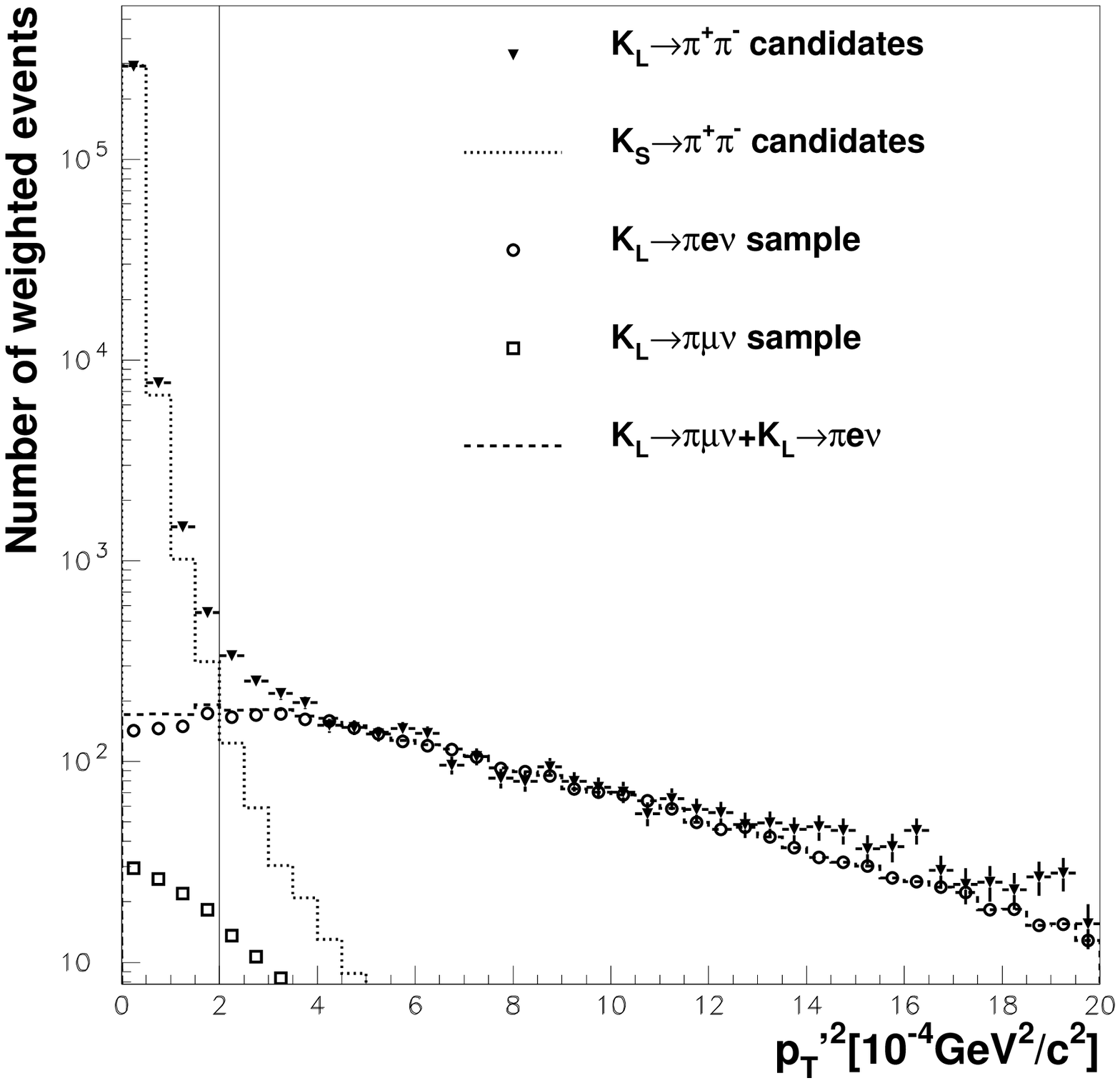}  
\caption{Distribution of $p_t{}'{}^2$ for $\kl$ candidates,
$\ks$ events (normalized in the first bin), and 
background from $\kethree$ and $\kmuthree$ decays.
The signal region is 
$p_t{}'{}^2$ $<$ $2\times 10^{-4}$ GeV$^2$/$c^2$.\label {fig:bkgc}}
\vspace{2cm}
\includegraphics*[scale=0.413]{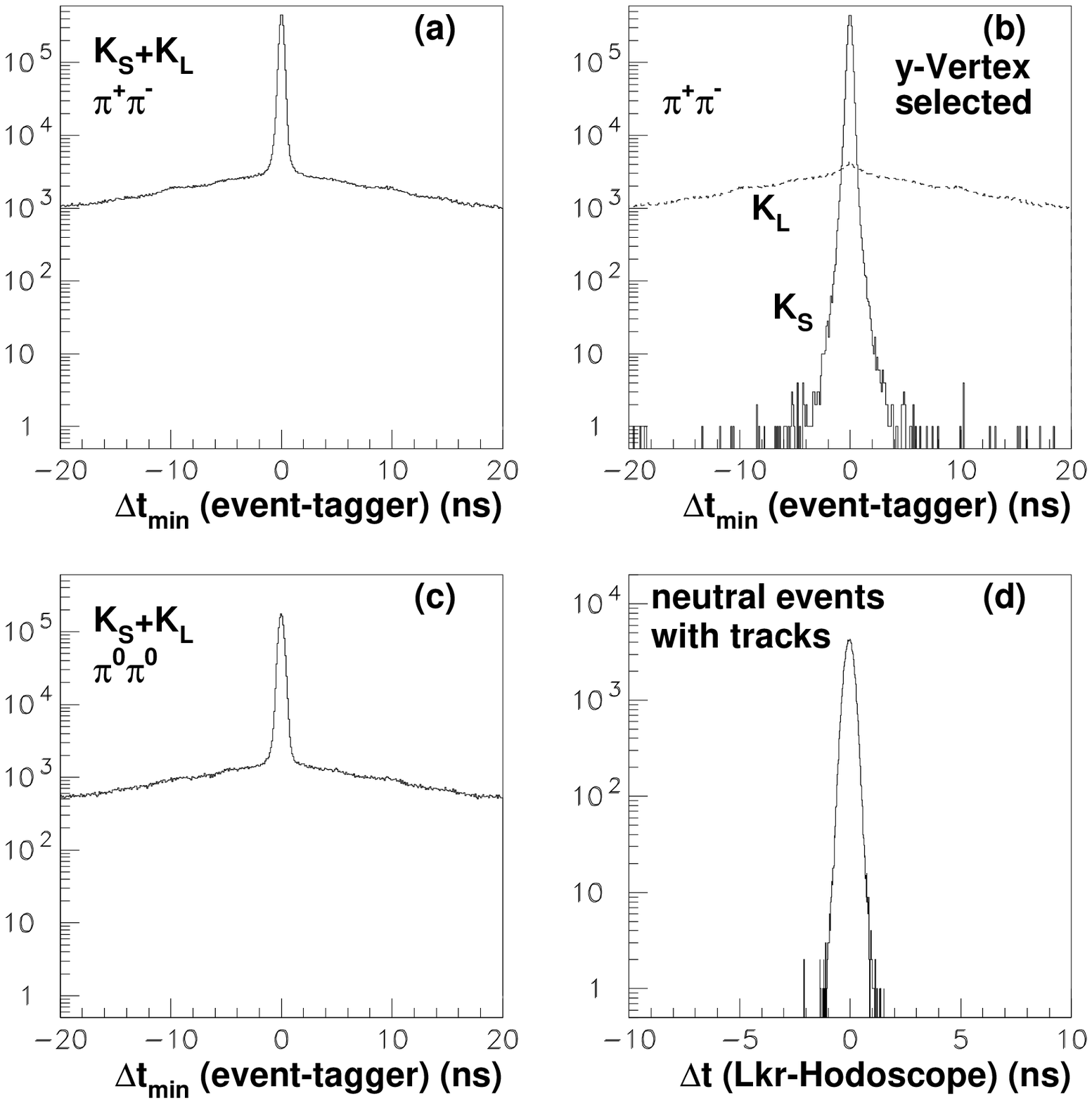}  
\caption {(a) Distribution of the minimum difference between
  tagger and event times for $\pipic$ decays.  The peak corresponds to
  $\ks$ events. (b) The same, separated into $\ks$ and $\kl$ events using the vertex
position. (c) as (a) for $\pipin$ mode.  (d) Coincidence time for
  neutral events with tracks.\label {fig:tagging}}
\end{center}
\end {figure}

\section{Event selection}
Decays to $\pipin$ are triggered in a 40~MHz pipeline,
which computes the number of clusters, the total energy and the first 
and second momenta of the energy distribution in the calorimeter. This allows
to select $\ko$ candidates with decay time $\tau\!<\!5\:\tau_S$. 
The trigger inefficiency is equal to $(12\pm4)\times 10^{-4}$. 
Off--line, the four clus\-ters are paired 
to check that their invariant 
mass is compatible with the $\pi^0$ mass (resolution 1.1~MeV/$c^2$). 
The corresponding $\chi^2$ distribution is shown in
figure~\ref{fig:bkgn}, where the difference between $\kl$ and $\ks$ is due to 
background from $3 \pi^0$ events, and is equal to $(8\pm 2)\times 10^{-4}$.

In $\pipic$ decays, a first level trigger 
is based on the scintillation hodoscope and the total energy measured by the 
calorimeters. This is down--scaled by two, 
and prompts the second level trigger, 
which uses drift chamber data to reconstruct vertices 
and compute invariant masses and decay times. 
The trigger efficiency is computed on samples of events 
from auxiliary, down--scaled triggers,
and found equal to $(91.0\pm0.1)$~\%.
The off--line se\-lec\-tion includes tighter cuts on invariant 
mass (2.5~MeV/$c^2$ resolution) and transverse mo\-men\-tum. 
The background from 
semileptonic $\kl$ decays is further rejected using the muon hodoscope, 
and the $E/P$ ratio from the e.m.\  calorimeter and the spectrometer. 
The total background in the
charged mode is $(23 \pm 4)\times 10^{-4}$ (see figure~\ref{fig:bkgc}).

Figure~\ref{fig:tagging} shows the distribution of the time difference 
between the tagger (on the $\ks$ proton beam line) 
and the main detector,
where the event time is measured by the e.m.\ calorimeter, 
and by the scintillation hodoscope, 
respectively for neutral and charged decays.
The coincidence window is $\pm\,2$~ns.
\begin{table}[th]
\begin{center}
\caption{Statistical samples, in thousands of events. \label{table:samples} }
\vspace{0.25cm}
\begin{tabular}{|c|c|c|c|} 
\hline
$\kl \rightarrow \pipin $ & 
$\ks \rightarrow \pipin $ & 
$\kl \rightarrow \pipic $ & 
$\ks \rightarrow \pipic $ \\
\hline
489 & 975 & 1,071 & 2,087 \\
\hline
\end{tabular}
\end{center}
\end{table}
For $\pipic$, the rate of inefficiency 
(resulting in a $\ks$ being misidentified as a $\kl$) 
and of accidental coincidences 
(tagging a $\kl$ as $\ks$) 
are measured applying a selection on the vertex position 
(figure~\ref{fig:tagging}b). 
In the neutral mode, accidental tagging is measured in off--set time intervals,
and the efficiency is obtained comparing the response of the calorimeter 
and the scintillation hodoscope in events with photon conversion or Dalitz decays
(figure~\ref{fig:tagging}d).
The double ratio $R\/$ is affected by the differences 
of the tagging errors between neutral and charged modes.  
The tagging inefficiency is about $1\times 10^{-4}$, 
with a difference of $(0 \pm 1)\times 10^{-4}$ 
between the two modes. 
The rate of accidental tag\-ging is 11.2~\% on average,
and is measured to be  
$(10 \pm 5)\times 10^{-4}$ larger in the neutral mode.

Table~\ref{table:samples} shows the number of events in each channel, after 
background and tagging corrections.

\section {Systematic corrections and uncertainties}
The difference between the decay distributions of
$\kl$ and $\ks$ in the 3.5 $\tau_S$ accepted interval implies a difference in
acceptance, which requires a correction to the measured value of $R\/$. 
This effect is reduced by 
weighting the $\kl$ events used in the double ratio with a function of the 
proper decay time $\tau$, proportional to the expected ratio of $\ks$ and $\kl$
decay rates. In the charged mode, residual differences in acceptance due to 
the beam geometries are minimized by a kinematical cut which rejects asymmetric
decays. Finally, to be independent of the $\pm\,10$~\%
difference in the energy spectra of the two beams, the analysis is performed 
in 5~GeV wide energy bins. The result of this procedure is an 
acceptance correction of $(29 \pm 12)\times 10^{-4}$, where most of the
uncertainty is due to Monte Carlo statistics.

The effects of accidental activity are minimized by taking data  
simultaneously in the four channels.
The rate of accidentals is measured to be equal for 
$\kl$ and $\kl$ events within the error of 1~\%.
\begin {figure}[h]
\begin{center}
\includegraphics*[scale=0.415]{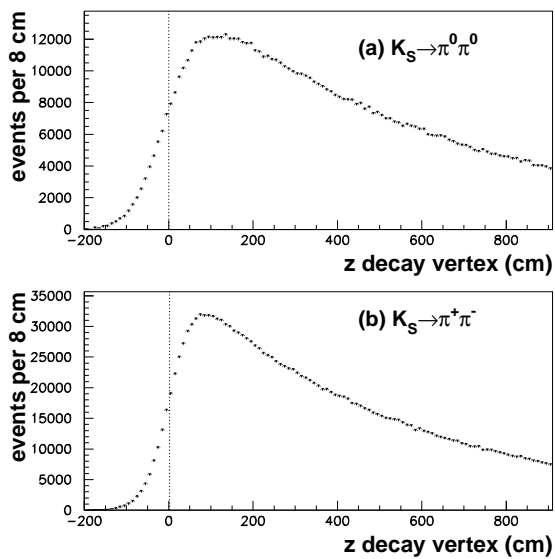}  
\caption{Distribution of the reconstructed decay vertex in $K_S$ events
for (a) the $\pipin$ mode and (b) the $\pipic$ mode. The
rising edge corresponds to the position of the veto--counter
(dotted line).}
\label{fig:aks}
\end{center}
\end {figure}
A small correction to $R\/$ 
is computed using a sample of
events overlayed with data selected by random triggers, 
hence obtaining a sample with artificially higher 
accidental activity. Another small correction is due to beam scattering in the 
collimators of the $\kl$ beam.

Particular care is taken in the definition of the accepted decay range.
For $\ks$ decays, the upstream end of the fiducial volume is defined by 
a veto--counter placed across the beam. 
The absolute energy scale of the e.m.\ calorimeter is tuned by 
checking the distribution of the reconstructed events against the position 
of the veto--counter (figure~\ref{fig:aks}). 
For $\kl$ decays, the definition of the boundaries 
$0\!<\!\tau\!<\!3.5\,\tau_S$ relies on an ac\-cu\-rate knowledge 
of the energy scale, and
of the linearity of the calorimeter response. Linearity and uniformity  
are studied with electrons from $K_{e3}$ events continuously 
recorded, and with auxiliary data. The task is facilitated by the intrinsic 
high stability of the detector (the absolute en\-er\-gy scale was stable within 
$\pm\, 5\times 10^{-4}$ throughout the entire run). 
The total contribution to the systematic
error in $R\/$ is $\pm\, 12\times 10^{-4}$.
 In the charged mode, the fit to the $\ks$ veto--counter provides 
a check on length scales and alignment, quantified in a small 
systematic uncertainty.

Table~\ref{table:corrections} lists all the corrections applied to the 
double ratio. The uncertainties in the first four lines are dominated by 
the statistics of the control sample used in each study.
\begin{table}
\begin{center}
\caption{Corrections and uncertainties to $R\/$, in $10^{-4}$ units. 
\label{table:corrections} }
\vspace{0.25cm}
\begin{tabular}{|l|crcr|} 
\hline
Tagging errors 	             & $+$ & 18 & $\pm$ & 11 \\
$\pipic$ trigger efficiency  & $+$ &  9 & $\pm$ & 23 \\
Acceptance                   & $+$ & 29 & $\pm$ & 12 \\
Accidental effects     	     & $-$ &  2 & $\pm$ & 14 \\
$\pipin$ background          & $-$ &  8 & $\pm$ &  2 \\
$\pipic$ background	     & $+$ & 23 & $\pm$ &  4 \\
Beam scattering              & $-$ & 12 & $\pm$ &  3 \\
Energy scale and linearity   &	 &    & $\pm$ & 12 \\
Charged vertex 	             &   &    & $\pm$ &  5 \\
\hline
Total correction             & $+$ & 57 & $\pm$ & 35 \\
 \hline
 \end{tabular}
 \end{center}
 \end{table}
\section {Result}
\begin {figure}[h]
\begin{center}
\includegraphics*[scale=0.415]{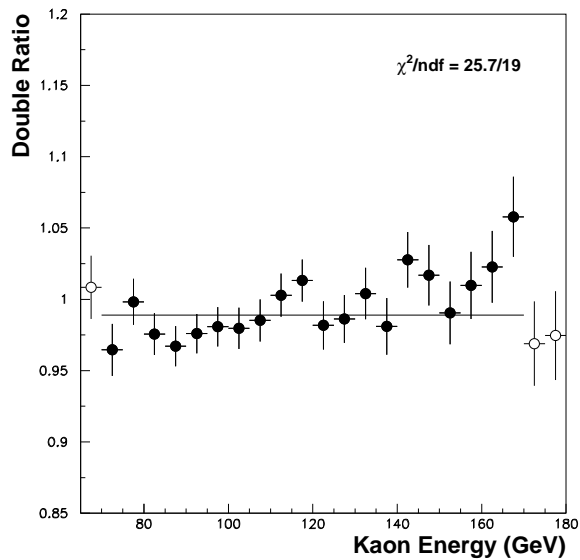}  
\caption{Measured double ratio in energy bins. The points used for the
measurement of $\repr$ are shown in black.}
\label {fig:r_e}
\end{center}
\end {figure}

Figure~\ref{fig:r_e} shows the result for $R\/$ in the different energy bins.
Corrections for trigger efficiency, tag\-ging, background
and acceptance are included in each bin.
The overall average is $R=0.9889\pm0.0027\pm0.0035$, where the first error
is from the statistical fluctuation in the event samples, and the second 
is from the uncertainties in table~\ref{table:corrections}.
The three blank points at the extremes were studied as an additional check on 
systematic effects, and including them would not modify significantly the
average value.
The corresponding result for the parameter describing direct CP violation is:
$$\repr = (18.5 \pm 4.5 \pm 5.8)\times 10^{-4}$$
or, combining the errors in quadrature, 
$\repr =  (18.5 \pm 7.3)\times 10^{-4}$.

\begin {figure}[h]
\begin{center}
\includegraphics*[scale=0.415]{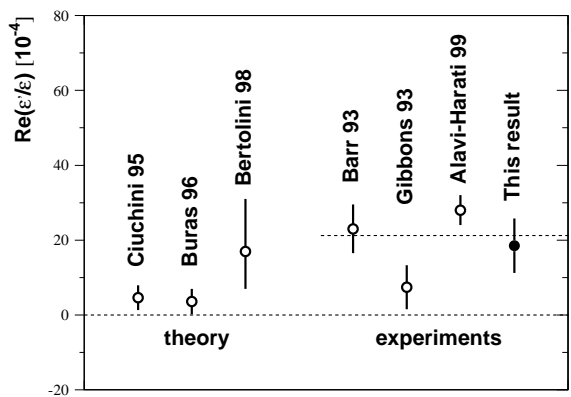}  
\caption{Comparison of prediction and experimental results.}
\label {fig:allresults}
\end{center}
\end {figure}
In figure~\ref{fig:allresults} this new result is compared with some 
predictions based on the Standard Model \cite{ref:computations} and other 
measurements \cite{ref:na31,ref:e731,ref:ktev}. 
The existence of direct CP violation in the neutral kaons
is confirmed at the level of $\repr \simeq  20\times 10^{-4}$.

A more accurate knowledge of $\repr$ will be possible with data 
collected by NA48 in 1998, 1999, and 2000. An increase of the statistical 
samples by a factor $\simeq \,10$ is expected, together with a significant 
reduction in the systematic uncertainties.

 \end{document}